\documentclass[aps,prl,reprint]{revtex4}

\usepackage{graphicx}
\usepackage{amsmath}
\usepackage{textcomp} 

\begin{document}


\title{Interference of coherent spin waves in micron-sized ferromagnetic waveguides}


\author{%
 Philipp Pirro\textsuperscript{\textsf{\bfseries }},
Thomas Br\"acher \textsuperscript{\textsf{\bfseries }},
Katrin Vogt\textsuperscript{\textsf{\bfseries }}, 
Bj\"orn Obry\textsuperscript{\textsf{\bfseries }},  
Helmut Schultheiss\textsuperscript{\textsf{\bfseries }}, 
Britta Leven\textsuperscript{\textsf{\bfseries }}, 
and Burkard Hillebrands\textsuperscript{\textsf{\bfseries }}
 }



\affiliation{Fachbereich Physik and Forschungszentrum OPTIMAS, Technische Universit\"at Kaiserslautern, D-67663 Kaiserslautern, Germany}



\newcommand{\units}[3][]{$#1\mathrm{#2\,#3}$}
\date{\today}
\begin{abstract}
We present experimental observations of the interference of spin-wave modes propagating in opposite directions in micron-sized Ni$_{81}$Fe$_{19}$-waveguides. To monitor the local spin-wave intensity distribution and phase of the formed interference pattern, we use Brillouin light scattering micro\-scopy. The two-dimensional spin-wave intensity map can be understood by considering the interference of several waveguide eigenmodes with different wavevectors quantized across the width of the stripe. The phase shows a transition from linear dependence on the space coordinate near the antennas characteristic for propagating waves to discrete values in the center region characteristic for standing waves. 
\end{abstract}
%
%
\maketitle   
  \begin{figure}[]
	 \centering
	  \includegraphics[width=0.45\textwidth]{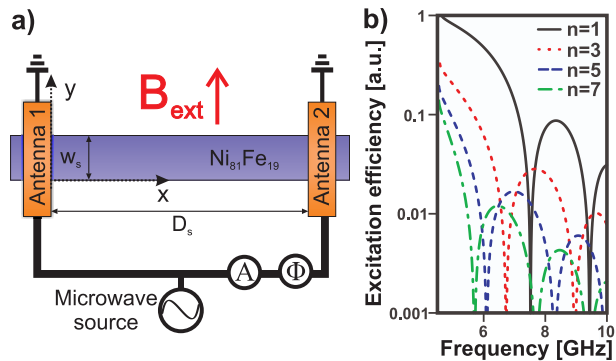}
	  \caption{\label{fig:Design}a) Schematic sample design: a Ni$_{81}$Fe$_{19}$ microstripe (width  \units[w_\mathrm{s}=]{4.1}{\mu m}, thickness \units[t=]{40}{nm}) is placed between two microwave antennas (width \units[w_\mathrm{a}=]{2.2}{\mu m}) which are connected to the same microwave source ($\sim$). A phase shifter ($\Phi$) and an attenuator (A) are used to shift the relative phase and the amplitude between the two antennas respectively. b)\,Calculated excitation efficiencies for the width modes $n=1,3,5,7$ (para\-meter see Fig.\,\ref{fig:DispersionTheo}) for the particular stripe and antennas shown in a) and an external magnetic field of \units[B_\mathrm{ext}=]{30}{mT}. Due to symmetry reasons, the excitation efficiency for the even modes vanishes.}
\end{figure}

\paragraph{Introduction}The investigation of spin waves in micron-sized metallic ferromagnetic structures has been the subject of several experimental \cite{Vogt2009,Demidov2009-1,Schultheiss2008,Chumak2009} studies due to their potential application in microwave signal processing and spin-wave logic devices \cite{Schneider2008-1,Khitun2008}, but also due to the possibility to address basic physical phenomena related to the spin transfer torque \cite{Vlaminck2008,Seo2009} and the interaction of spin waves with topological objects like domain walls \cite{Hermsdoerfer2009,Bayer2005,Hertel2004,LeMaho2009}. In all experimental studies, understanding the excitation and propagation of spin waves is of vital importance, especially for some concepts of next-generation logic circuits which use the phase and amplitude of spin waves as an information carrier. As the processing of data in these devices is performed by the interference of different input waves, the ability of the spin waves to form stable and predictable interference patterns is of crucial importance. An issue of interest here is the determination of the length scale over which spin waves can propagate without loss of coherence. Several numerical studies \cite{Lee2008,Choi2006} were dedicated to the interference of spin waves in micro\-structures but experimental realizations are still limited (see e.g. \cite{Perzlmaier2008}).

In this letter, we report on the experimental observation of interfering spin waves in micron-sized waveguides of Ni$_{81}$Fe$_{19}$ (see Fig.\,\ref{fig:Design}a for detailed sample design). We find that the coherence length is at least as long as the achieved propagation distance of the locally excited spin waves and is of the order of several microns. In addition, we show that the complexity of the formed spin-wave intensity patterns can be predicted quantitatively.
\paragraph{Sample design} The investigated Ni$_{81}$Fe$_{19}$ structures have a thickness of \units[t=]{40}{nm}, a width $w_\mathrm{s}$ of \units{4.1}{\mu m} and lengths in the range of 15 to 20\,$\mu m$ and are patterned using electron beam lithography and lift-off technique. On top of the Ni$_{81}$Fe$_{19}$ structures, two antennas (Cu) are processed which are connected to the same microwave source to excite phase locked spin waves at $x=0$ and $x=D_\mathrm{s}$. These waves propagate in opposite directions towards the center of the structure and interfere. The width of the antennas is  \units[w_\mathrm{a}=]{2.2}{\mu m}, resulting in efficient excitation of spin waves with wavelengths in the micron range. 
\paragraph{Experimental results}
To observe the spin-wave interference experimentally, we employ Brillouin light scattering (BLS) microscopy (for details see \cite{Vogt2009,Hillebrands}). Spin waves are excited concurrently with both antennas at a monochromatic frequency and the space-resolved BLS intensity (which is proportional to the spin-wave intensity) is recorded. During our measurements we apply a magnetic field $B_\mathrm{ext}$ in the range from 30 to 40\,mT  to align the magnetic moments perpendicular to the length of the stripe (y-direction).
\begin{figure}[h]
	 \centering
	  \includegraphics[width=0.5\textwidth]{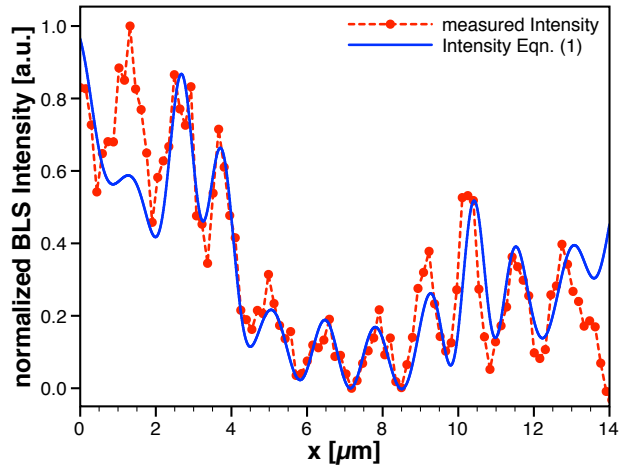}
	  \caption{\label{fig:Interferenz1}Red dots: BLS intensity of interfering spin waves with a frequency \units[\nu=]{7.13}{GHz} at a field \units[B_\mathrm{ext}=]{30}{mT} and an antenna spacing $D_\mathrm{s} \approx 14.5~\mu$m. Blue dotted line: Spin-wave intensity calculated according to Eqn. (\ref{Inten}) taking into account transversal modes up to $n=7$ (along $x$-axis, at $y=w_\mathrm{s}/2$).  }
\end{figure}

\begin{figure}[h]
	  \includegraphics[width=0.5\textwidth]{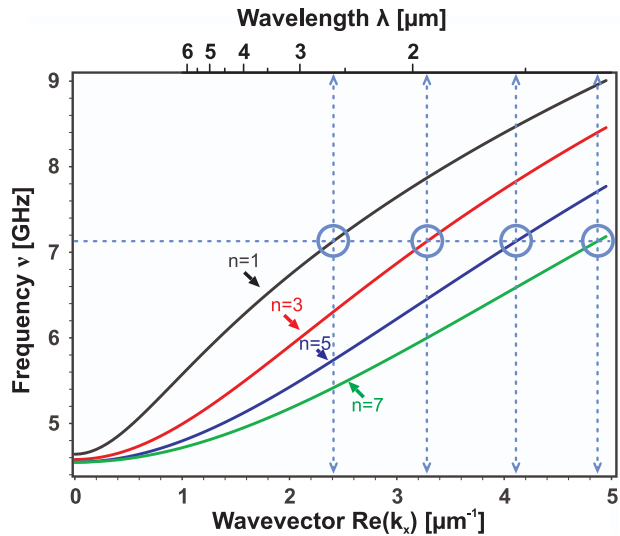}
	  \caption{\label{fig:DispersionTheo}Calculated dispersion relations for a Ni$_{81}$Fe$_{19}$-microstripe (\units[w_\mathrm{eff}=]{3}{\mu m}, thickness \units[t=]{40}{nm}, \units[B_\mathrm{ext}=]{30}{mT}) for the transversal modes $n=1,3,5,7$. Circles and dotted lines indicate the resulting wavelengths and wavevectors at the used frequency of \units[\nu=]{7.13}{GHz}.}
\end{figure}

A typical measurement taken in the middle of the stripe along the x-axis is shown in Fig.\,\ref{fig:Interferenz1}. Here, the stationary intensity maxima and minima that are expected in the case of  two phase locked waves propagating in opposite directions are clearly visible. 
To compare our results with the theo\-retical predictions, the dispersion relations for different transversal modes \cite{footnote} of mode order $n$ are shown in Fig.\,\ref{fig:DispersionTheo}. They are calculated according to \cite{Kalinikos1986} with the material parameters as summarized in \cite{Parameter}. Modes with even $n$ cannot be excited with these antennas because the excitation field is homogenous along the $y$ direction and these modes have no fluctuating magnetic moment averaged across the stipe width \cite{Demidov2009-2}.
The distances between the measured intensity maxima in Fig.\,\ref{fig:Interferenz1} show a good agreement with half of the wavelength of the first spin-wave mode $n=1$ in Fig.\,\ref{fig:DispersionTheo}. 

Nevertheless, to understand the spacial intensity distribution, especially effects like the increase of intensity for the maximum at $x=10.5~\mu m$, the interference of the spin waves needs to be described in more detail by taking into account higher transversal modes also. 
We model the intensity of the spin waves assuming sinusoidal transverse mode profiles in $y$ direction. The intensity of the formed stationary spin-wave interference pattern (averaged over one oscillation period) in between the antennas is then described by:
\begin{eqnarray}
\label{Inten}
I (\nu)&=& |A|^2\nonumber\\
&=&\biggl | \sum_{n} [a_n \sin \left ( \frac{n \pi }{w_{eff}}y\right) e^{i(k_{x}^{n}(\nu) x+\Phi)}  \nonumber \\
 &+& b \cdot  a_n \sin \left ( \frac{ n \pi}{w_{eff}}y\right ) e^{-i(k_{x}^{n}(\nu) (D_s-x))}]\biggl | ^2 \\ \nonumber
\end{eqnarray}
where $I$ denotes the spin-wave intensity, $n$ the mode number with respect to the transversal quantization, $a_n$ the amplitudes of the waves traveling in $+x$ direction, b the factor of imbalance of the antennas \cite{footnote2} (thus $a_n \cdot b$ are the amplitudes of the waves traveling in $-x$ direction) and $D_\mathrm{s}$ the distance between the antennas.
The complex wavevector $k_{\mathrm{x}}^{n}$ is related to the exponential amplitude decay length $\delta^{n}$ and the wavelength $\lambda^{n}$ via:
\begin{equation}
\label{k}
\operatorname{Im}(k_{\mathrm{x}}^{n})=\frac{1}{\delta^{n}}\ ; \ \operatorname{Re}(k_{\mathrm{x}})=\frac{2 \pi}{\lambda^{n}}.
\end{equation}
The decay length $\delta^{n}$ was determined using the group velocity $v_{\mathrm{g}}^n$ and the lifetime $\tau$ of the spin waves: 
\begin{equation}
\label{delta}
\delta^{n}=v_{\mathrm{g}}^n \tau=\frac{\partial \omega^n}{\partial k^{n}_{\mathrm{x}}} \tau .
\end{equation}
We extracted $v_{\mathrmÊ{g}}^n$ for the different modes from the dispersion relations shown in Fig.\,\ref{fig:DispersionTheo}. $\tau$ was calculated according to \cite{Stancil}, assuming the standard damping parameter $\alpha$ for Ni$_{81}$Fe$_{19}$ ($\alpha=0.01$). Since we computed the frequency dependent relative excitation amplitudes $a_1 / a_n$  for the different width modes $n$ (see Fig.\,\ref{fig:Design}b) by using the theory presented in \cite{Demidov2009-2,SchneiderReciprocity} the only fitted parameters are the relative phase shift $\Phi$ between the two antennas and the factor of imbalance $b$, which both strongly depend on the experimental conditions (cable length, contacts).

As one can see in Fig.\,\ref{fig:Interferenz1}, there is a good agreement between the measured intensity and the calculation using Eqn.\,(\ref{Inten}), proving that we can predict all physical relevant parameters like wavelength, group velocity and relative excitation efficiency. In detail, only modes with uneven $n$ can be excited by the antennas and only modes with $n \le 7$ contribute significantly. The group velocity $v_{\mathrm{g}}^n$ (and consequently the decay length $\delta^{n}$) decreases with increasing $n$. In the case of the chosen frequency \units[\nu=]{7.13}{GHz}, this is also true for the relative amplitudes $a_n$ (see Fig.\,\ref{fig:Design}b). The propagating spin waves show a constant degree of coherence, even though their amplitudes are strongly damped (for the first mode $\units[$\delta^{1}=]{5.3}{\mu m}). Thus, we can conclude that the length scale over which spin waves can propagate without losing their coherence is higher than the exponential amplitude decay length $\delta$.

\begin{figure}[]
	 \centering
	  \includegraphics[width=0.45\textwidth]{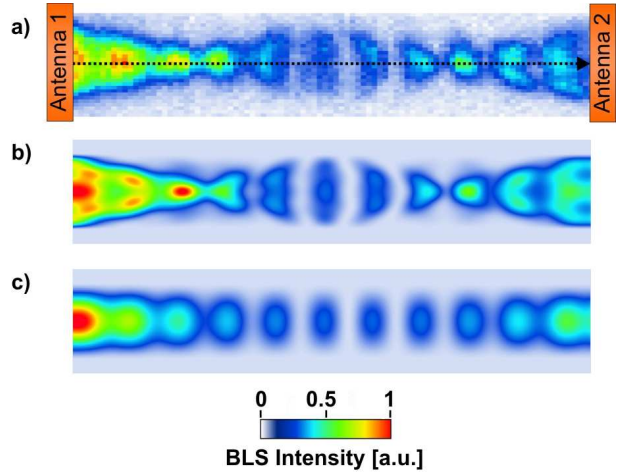}
	  \caption{\label{fig:2Dmap}a) Two-dimensional BLS-intensity map showing the interference of propagating spin waves in a Ni$_{81}$Fe$_{19}$-stripe (see Fig.\,\ref{fig:Design} and Fig.\,\ref{fig:Interferenz1} for parameters). b) Calculations taking into account the modes \mbox{$n=1,3,5,7$}. c) Calculations taking into account only the first mode $n=1$.}
\end{figure}

To visualize the influence of the different transversal modes, it is convenient to look at the two-dimensional spin-wave intensity shown in Fig.\,\ref{fig:2Dmap}, where the BLS measurement a) is compared to the complete calculation of Eqn.\,(\ref{Inten}) with $n=1,3,5,7$ in b) and to the calculation taking into account only the first mode $n=1$ in c). It is obvious that the interference of the different width modes leads to periodical convergence of the intensity (see \cite{Buettner1998}). This explains the local maximum shown in Fig.\,\ref{fig:Interferenz1} (at $x=10.5~\mu m$): it is situated at a point where the spin-wave intensity is concentrated in the middle of the stripe.
\begin{figure}[b]
	 \centering
	  \includegraphics[width=0.45\textwidth]{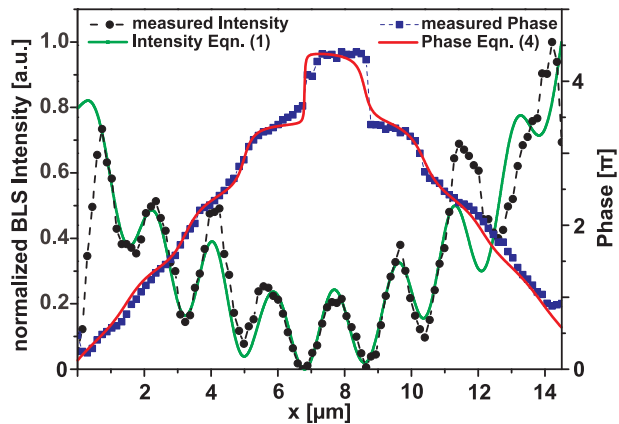}
	  \caption{\label{fig:Phase}Blue squares: Phase of interfering spin waves (at \units[B_\mathrm{ext}=]{40}{mT}, \units[\nu=]{7.13}{GHz}) measured with phase-resolved BLS microscopy and the comparison to the calculated phase according to Eqn.~(\ref{Phase}) (red line). Black dots:  Corresponding BLS intensity and its calculation with Eqn.~(\ref{Inten}) where the same parameters as for the phase calculation are used.  }
\end{figure}

To further verify our results, we use phase-resolved Brillouin light scattering microscopy \cite{Vogt2009,Demidov2009-1} to directly access the phase of the interfering spin waves. In Fig.\,\ref{fig:Phase}, the measured phase $\varphi$  is compared to the predicted values of the model in Eqn.\,(\ref{Inten}):
\begin{equation}
\label{Phase}
 \varphi=\arctan \left (\frac{\operatorname{Im}(A)}{\operatorname{Re}(A)}\right )
\end{equation}
where the complex amplitude $A$ of the spin waves is calculated according to Eqn.\,(\ref{Inten}). The results show an excellent agreement with the theoretical predictions for this measurement as well. 

As can be seen in Fig.\,\ref{fig:Phase} one can nicely distinguish two regions: Near the antennas, the phase is dominated by one of the propagating waves, hence a linear increase in phase is visible. In the middle between both antennas, both waves have comparable intensities and form a standing wave. The phase profile of standing spin waves consists of regions of constant phase which range over half the spin-wave wavelength $\lambda$ and are separated by phase jumps of $\pi$. Two of these phase jumps can be seen in Fig.\,\ref{fig:Phase}. The plateau in the center has a width \units[\Delta x \approx]{1.8}{\mu m} which corresponds well to the value of $\lambda/2$ of the dominating first transversal mode.

\begin{figure}[h]
	 \centering
	  \includegraphics[width=0.45\textwidth]{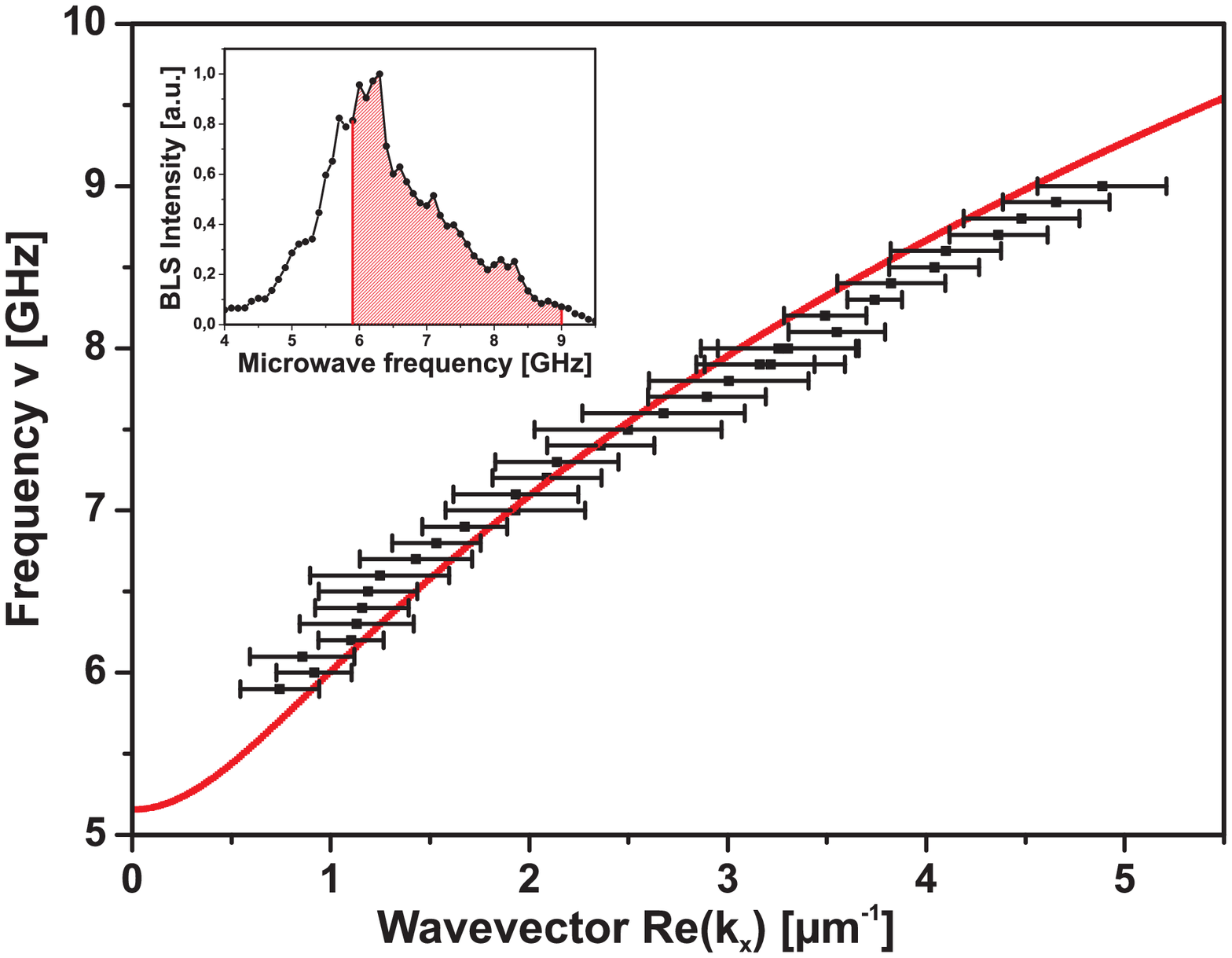}
	  \caption{\label{fig:Dispersion}Dispersion relation for the first width mode of the Ni$_{81}$Fe$_{19}$-stripe at \units[B_\mathrm{ext}=]{35}{mT}. The values for Re($k_{x}$) are measured via the average distance between two intensity maxima (see text). The inset shows the BLS intensity as function of the applied microwave excitation frequency. The marked area indicates the range in which Re($k_{x}$) was determined.}
\end{figure}

As it was already mentioned above, the spin waves are found to interfere coherently whenever a significant intensity from both antennas can be detected. Consequently, we can also employ the interference of the spin waves to verify the validity of the theoretical model used to calculate the dispersion relations in Eqn.\,(\ref{Inten}). To do this, we first measure the BLS intensity at a fixed position as a function of the microwave frequency (see inset of Fig.\,\ref{fig:Dispersion}) to identify the frequency range in which propagating spin waves can be excited. Then, space-resolved scans similar to the one in Fig.\,\ref{fig:Interferenz1} are performed for different excitation frequencies in this range.

We take the average distance $\Delta$ between two intensity maxima to calculate the wavevector of the first transversal mode: $k_\mathrm{x}^1=\pi / \Delta$ (see Fig~\ref{fig:Dispersion}). Thus, the influence of higher transversal modes on the distance between two maxima is neglected, which is a passable approximation for the par\-ticular experimental parameters (note that also in Fig.\,\ref{fig:2Dmap}, the higher transversal modes do not significantly change the distance between the maxima in the middle of the stripe). The remaining influence of the higher transversal modes leads to small variations of $\Delta$ around the mean value re\-presented by the error bars. The good agreement between the experimental results and theoretically calculated dispersion relation for the first transversal mode further justifies our approximation.

\paragraph{Conclusions}
We showed that propagating spin waves in microscopic waveguides interfere coherently, so the construction of micron-sized spin-wave logic devices should be principally possible. In addition, we demonstrated that wavelength, group velocity, decay length and excitation efficiency for the different transversal modes can be predicted by the theory for spin waves in thin magnetic films. Furthermore, the interference of the propagating spin waves can be used to measure their dispersion relation in an extended frequency range.
\begin{acknowledgments}
The authors thank Dr. A. Beck for deposition of the magnetic thin film and the Nano+Bio Center of the Technische Universit\"at Kaiserslautern for assistance in sample preparation. K. Vogt acknowledges financial support by the Carl-Zeiss-Stiftung. B. Obry acknowledges the Deutsche Forschungsgemeinschaft (Graduiertenkolleg 792) for financial support.
\end{acknowledgments}

%

\begin{thebibliography}{24}
\bibitem{footnote}
Due to the confinement of the wave in y-direction, $k_{y}$ is quantized: $k_{y} =\frac{n\cdot \pi}{w_{\mathrm{eff}}}$, $n \in 1,2,3,...$. $w_{\mathrm{eff}}$ is the effective quantization width of the spin wave (see \cite{BayerCrowell} for details). In our case, $w_\mathrm{eff}$ was approximated by the measured width of the transverse intensity profile ($\approx\,3~\mu$m for the applied magnetic fields).
\end{thebibliography}
%

\end{document}